\documentclass[]{aastex631}

\shorttitle{Lambda-CDM-NG model}
\shortauthors{Cook}

\graphicspath{{./}{figures/}}
\begin{document}

\title{The $\Lambda$CDM-NG cosmological model: A possible resolution of the Hubble tension}

\author[0000-0002-6243-8249]{Richard J. Cook}
\affiliation{U.S. Air Force Academy, CO 80840\footnote{Correspondence welcomed at:  richard.cook43@gmail.com} }
\begin{abstract}

We offer a cosmological model based on conventional general relativity (no speculative physics) which may will resolve the Hubble tension.  A reanalysis of the foundation of the Lambda-CDM model shows that general relativity alone does not specify what fraction of the mass density acts as the source term in Friedmann's equation and what fraction acts as the source of the gravitational potential of condensed objects. This observation opens the way to alternative cosmological models within conventional general relativity, and proves that the $\Lambda$CDM model is not the unique solution of Einstein's equations for the usual cosmological sources of gravitation. We emphasize that the source of the gravitation potential in the $\Lambda$CDM model is the deviation $\delta\rho_{m}$ of the mass density away from its average value, and not the total density of condensed masses as in Newtonian theory. Though not often stated, this is a modification of Newtonian gravitation within the $\Lambda$CDM model. The $\Lambda$CDM-NG model uses the freedom to move matter between source terms to restore the source of gravitational potential to its Newtonian form. There is no Hubble tension in the $\Lambda$CDM-NG model if the gravitational potential of condensed objects (stars, galaxies, and dark matter clouds) falls in a certain range, a range which does not seem unreasonable for the actual universes. The deceleration parameter in the $\Lambda$CDM-NG model differs from that in the $\Lambda$CDM model, suggesting a test to distinguish between the two models.
\end{abstract}

\keywords{Cosmology (343) --- Cosmological models (337) --- Relativistic cosmology (1387) ---Observational cosmology (1146)}

%********************************
\section{Introduction} \label{sec:intro}
%*******************************
Modern Cosmology began with Einstein's application of his newly minted General Theory of Relativity \citep{Einstein1} to the universe at large \citep{Einstein2}. Although this initial effort was flawed by Einstein's belief in a static universe, it was motivation for additional work by De Sitter \citep{De Sitter1, De Sitter2}, Friedman \citep{Friedmann1,Friedmann2}, Lema\^itre \citep{Lemaitre1,Lemaitre2}, Robertson \citep{Robertson1,Robertson2,Robertson3}, Walker \citep{Walker}, Eddington \citep{Eddington}, and others, showing that the universe must be dynamic. Then, with the discovery of dark matter \citep{Zwicky1,Zwicky2,Rubin,Rogstad} and the accelerated expansion of the universe \citep{Perlmutter,Riess}, cosmological modeling culminated in the current Lambda-CDM (or $\Lambda$CDM) model of the universe \citep{Weinberg}.

 At the present time the $\Lambda$CDM model is being challenged on three fronts: (1) The long standing cosmological constant problem \citep{Weinberg2}, (2) Initial observations by the Jame Webb Space Telescope (JWST); Advanced Deep Extragalactic Survey (JADES) \citep{Finkelstein,Labbe,Adams,Rodighiero} suggest galaxy formation at times earlier than expected from the $\Lambda$CDM model \citep{Boylan-Kolchin}; and (3) Measurements of the Hubble constant $H_{0}$ using ``late universe" (redshift $z\approx$ 0-1) Type Ia supernovae  give a value [$\sim 74\  km\, s^{-1}Mpc^{-1}$] significantly different from the value [$\sim 68\ km\,s^{-1}Mpc^{-1}$] obtained from observations of the ``early universe" (redshift $z=$ 1100) Cosmic Microwave Background (CMB) radiation  \citep{Planck}. The cosmological model proposed here makes a contributions to the last of these quandaries. 
 
The present paper in based on the insight, explained in Sec.2, that general relativity alone does not specify how much of the mass-energy of the universe serves as source term $\rho_{F}$ in Friedmann's equation and how much $\rho_{\Phi}$ is the source of gravitational potential $\Phi$, with different splittings of the total energy among the two source terms giving different cosmological models. Section 3 identifies the sources $\rho_{F}$ and $\rho_{\Phi}$ which give the $\Lambda$CDM model, and points out that the source of gravitational potential in this model is not the same as in Newtonian gravitation. The $\Lambda$CDM-NG model is introduced in Sec.4 as the cosmological model which restores the source of the gravitational potential to its Newtonian form. Sections 5-10 develop the formalism of the $\Lambda$CDM-NG model. Section 11 shows there is no Hubble tension in the $\Lambda$CDM-NG model, if the gravitational potential of the universe at the present time falls in a given range. Section 12 calculates the deceleration parameter for the $\Lambda$CDM-NG model, and points to the difference between this and that of the $\Lambda$CDM model as the basis of a possible test to distinguish between the two models. Several expansion histories, within the $\Lambda$CDM-NG model, illustrate, in Sec.13, the variety of solutions which have the observed values of cosmological parameters $\Omega_{\Lambda}$, $\Omega_{m}$, $\Omega_{r}$, $H_{0}$, and $q_{0}$; and at the same time exhibit no Hubble tension in this model. A physical interpretation of the $\Lambda$CDM-NG model is given in Sec.14 in term of gravitational time dilation of the standard comoving clocks which measure cosmic time. The paper concludes in Sec.15 with some final remarks.

%*************************
\section{Non-Uniqueness of the Lambda-CDM Model}  
%************************
From the time of Friedmann onward, cosmological models, including the $\Lambda$CDM model, have taken the Friedmann-Lema\^itre-Robertson-Walker (FLRW) line element as background geometry, and have assumed a source term in Friedmann's equation equal to the space-averaged density of \textit{all} mass-energy in the universe \citep{Friedmann1,Lemaitre1,Eddington,Weinberg}. With the mean value $\bar{\rho}(t)$ of the inhomogeneous density $\rho(\mathbf{x},t)$ of the universe taken as the source term in Friedmann's equation, there remains only the deviation $\delta\rho=\rho(\mathbf{x},t)-\bar{\rho}(t)$ of $\rho(\mathbf{x},t)$ away from its average value $\bar{\rho}(t)$ as the source of gravitational potential $\Phi(\mathbf{x},t)$. In this section, and the next two, we show that this very plausible approach to cosmological theory has two consequences: (1) it obscures the existence of alternative cosmological models within conventional general relativity, and with the same sources of gravitation as the $\Lambda$CDM model; and (2) it necessarily involves a modification of Newtonian gravitation where Newtonian theory is expected to be valid in general relativity.

The non-uniqueness of the $\Lambda$CDM model is easily shown if we refrain from specifying the source term in Friedmann's equation \textit{a priori}, and instead consider, from the beginning, a line element  of sufficient generality to describe both the large-scale expansion of the universe, as described by the scale factor $a(t)$, and the smaller scale perturbations of the metric by the inhomogeneous density $\rho(\mathbf{x},t)$. It is usual in structure formation calculations to write such a line element as
\begin{equation}
ds^2=-\left(1+\frac{2\Phi}{c^2}\right)c^2dt^{\,2}+a^{2}(\tilde{t})\left(1-\frac{2\Psi}{c^2}\right)\delta_{ij}dx^{i} dx^{j},
\label{ScalarPert}
\end{equation}
(often with different sign conventions and labels $\Phi$ and $\Psi$ switched). This is the flat space LFRW line element perturbed by scalar functions $\Phi(\tilde{\mathbf{x}},\tilde{t})$ and $\Psi(\tilde{\mathbf{x}},\tilde{t})$ (gauge invariant Bardeen potentials \citep{Bardeen}) in what is called the \textit{conformal Newtonian gauge}. If we are willing to ignore certain small effects due to neutrino anisotropic stresses during the early radiation-dominated era, in order to simplify the following argument, then $\Psi$ and $\Phi$ are equal, and line element (\ref{ScalarPert}) becomes \citep{Dodelson}
\begin{equation}
ds^2=-\left(1+\frac{2\Phi}{c^2}\right)c^2dt^2+a^{2}(\tilde{t})\left(1-\frac{2\Phi}{c^2}\right)\delta_{ij}dx^{i} dx^{j}.
\label{ScalarPe}
\end{equation}
where $\Phi$ is recognized as the Newtonian gravitational potential. This then is the simplest line element with sufficient detail to make the following argument. It contains the scale factor $a(t)$ describing the large-scale dynamics of the universe and the gravitational potential $\Phi(\mathbf{x},t)$ describing smaller scale gravitational interactions. 

Line element (\ref{ScalarPe}) is usually applied in situations where $\Phi/c^2$ is small ($\Phi/c^2<<1$) and $\Phi/c^2$ constitutes a weak perturbation to the unperturbed line element
\begin{equation}
ds^2=-c^2dt^2+a^{2}(t)\delta_{ij}dx^{i} dx^{j}.
\label{FriedUnPur}
\end{equation}
We, however, because we are interested in late times when $\Phi/c^2$ is not necessarily small, shall treat Eq.(\ref{ScalarPe}) as an exact line element without the restriction that $\Phi/c^2$ be small. The exact Einstein equation
\begin{equation}
 G^{0}_{\ 0}=(8\pi G/c^4)T^{0}_{\ \,0}
 \label{Ein00}
 \end{equation}
for metric (\ref{ScalarPe}), with energy-momentum-tensor component $T^{0}_{\ \,0}=-\rho c^2$, reads. 
\begin{equation}
\frac{1}{1+2\Phi/c^2}\left[\frac{1}{a}\frac{da}{dt}-\frac{1}{1-2\Phi/c^2}\frac{\partial}{\partial t}\left(\frac{\Phi}{c^2}\right)\right]^2+\frac{1}{(1-2\Phi/c^2)^2}\left[\frac{2}{3}\frac{\mathbf{\nabla}^2\Phi}{a^2}+\frac{\mathbf{\nabla}\Phi\mathbf{\cdot}\mathbf{\nabla}\Phi}{c^2a^2 (1-2\Phi/c^2)}\right]=\frac{8\pi G}{3}(\rho_{\Lambda}+\rho_{m}+\rho_{r}),
\label{FullEin00}
\end{equation}
where $(\mathbf{\nabla}\Phi)_{i}=\partial\Phi/\partial x^{i}$,  $\mathbf{\nabla}^2\Phi=\delta^{ij}\partial^2\Phi/\partial x^{i}\partial x^{j}$, and we have written the density
\begin{equation}
\rho(\mathbf{x},t)=\rho_{\Lambda}+\rho_{r}(t)+\rho_{m}(\mathbf{x},t),
\label{TrueDens}
\end{equation}
as the sum of a perfectly uniform dark energy, $\rho_{\Lambda}$= constant, homogeneous radiation, $\rho_{r}(t)= \rho_{r}(t_{0})/a^{4}(t)$, and an inhomogeneous matter density, $\rho_{m}(\mathbf{x},t)$, describing the combined densities of dark matter and baryonic matter (stars, galaxies, and dark matter clouds). 

We learn from Eq.(\ref{FullEin00}) the simple, but important, fact that general relativity does not specify how the various densities on the right in this equation are partitioned into source terms for the various field terms on the left side of the equation. Any reasonable association of densities with field terms gives a solution of Einstein equation (\ref{Ein00}).

To be clear, we split the total density $\rho$ into a source term $\rho_{F}$ for what will be the Friedmann equation, and a source term $\rho_{\Phi}$ for what will be the equation for potential $\Phi$. We argue that, if the source term $\rho_{\Phi}$ is null, the potential $\Phi$ will also vanish. Using these values in (\ref{FullEin00}), we obtain the familiar Friedmann equation 
\begin{equation}
\left(\frac{1}{a}\frac{da}{dt}\right)^2=\frac{8\pi G}{3}\rho_{F},
\label{FriedGot}
\end{equation}
and subtracting this from Eq.(\ref{FullEin00}), we have the equation for potential $\Phi$, 
\begin{equation}
\frac{1}{1+2\Phi/c^2}\left[\frac{1}{a}\frac{da}{dt}-\frac{1}{1-2\Phi/c^2}\frac{\partial}{\partial t}\left(\frac{\Phi}{c^2}\right)\right]^2-\left(\frac{1}{a}\frac{da}{dt}\right)^2+\frac{1}{(1-2\Phi/c^2)^2}\left[\frac{2}{3}\frac{\mathbf{\nabla}^2\Phi}{a^2}+\frac{\mathbf{\nabla}\Phi\mathbf{\cdot}\mathbf{\nabla}\Phi}{c^2a^2 (1-2\Phi/c^2)}\right]=\frac{8\pi G}{3}\rho_{\Phi}.
\label{PotEqGen}
\end{equation}
As a check on this result, we evaluate Eq.(\ref{PotEqGen}) to first order in the dimensionless potential $\Phi/c^2$, and obtain the linearized potential equation familiar from structure formation theory
\begin{equation}
\frac{\mathbf{\nabla}^2\Phi}{a^2} -3\frac{\Phi}{c^2}\left(\frac{1}{a}\frac{da}{dt}\right)^2-3\left(\frac{1}{a}\frac{da}{dt}\right)\,\frac{\partial}{\partial t}\left(\frac{\Phi}{c^2}\right)=4\pi G \rho_{\Phi},
\label{PotEq}
\end{equation}
except that the source $\delta\rho_{m}$ is replaced by the density $\rho_{\Phi}$ \citep{Baumann}.

The sum of $\rho_{F}$ and $\rho_{\Phi}$ is, of course, the total density,
\begin{equation}
\rho_{\Phi}+\rho_{F}=\rho_{\Lambda}+\rho_{r}(t)+\rho_{m}(\mathbf{x},t).
\label{Constraint}
\end{equation}
It is important to note that, according to this equation, whatever density $\rho_{\Phi}$ is chosen as the source term for the gravitational potential, Eq.(\ref{PotEqGen}), this density is unavailable to influence the evolution of the scale factor $a(t)$ as part of the source term in Friedmann's equation (\ref{FriedGot}).

The all-important point of this section is that different partitions of the total density into source terms $\rho_{F}$ and $\rho_{\Phi}$ for Eqs.(\ref{FriedGot}) and (\ref{PotEqGen}) give different cosmological models. One partition gives the $\Lambda$CDM model and a different partition gives the $\Lambda$CDM-NG model. We emphasize that the splitting of the total density into $\rho_{\Phi}$ and $\rho_{F}$ is a choice made outside of general relativity. It is an independent assumption, or hypothesis, requiring its own justification.

%***********************
\section{The Lambda-CDM Model}
%**********************
The $\Lambda$CDM model is based on the assumption that the source term in Friedmann's equation is the space-averaged density of all mass-energy in the universe,
\begin{equation}
\rho_{F}(t)=<\rho(\mathbf{x},t)>=\rho_{\Lambda}+\rho_{r}(t)+\bar{\rho}_{m}(t),
\label{SourceF}
\end{equation}
where $\bar{\rho}_{m}=<\rho_{m}(\mathbf{x},t)>$. Then, according to Eq.(\ref{Constraint}), the source of the gravitational potential $\Phi$ is necessarily 
\begin{equation}
\rho_{\Phi}(\mathbf{x},t)=\delta\rho_{m}=\rho_{m}(\mathbf{x},t)-\bar{\rho}_{m}(t),
\label{SourceN}
\end{equation}
which is the deviation of $\rho_{m}(\mathbf{x},t)$ away from its average value (Fluctuations of radiation also contribute to the source of gravitational potential, but for simplicity of presentation we consider only the dominant matter fluctuations).

One version of the basic equations of the $\Lambda$CDM model (neglecting neutrino anisotropic stresses) is written as
\begin{equation}
ds^2=-\left(1+\frac{2\Phi}{c^2}\right)c^2dt^2+a^{2}(t)\left(1-\frac{2\Phi}{c^2}\right)\delta_{ij}dx^{i} dx^{j},
\label{ScalarPe2}
\end{equation}
\begin{equation}
\left(\frac{1}{a}\frac{da}{dt}\right)^2=\frac{8\pi G}{3}\left[\rho_{\Lambda}+\rho_{r}(t)+\bar{\rho}_{m}(t)\right],
\label{Friedmann2}
\end{equation}
and
\begin{equation}
\frac{\mathbf{\nabla}^2\Phi}{a^2} -3\frac{\Phi}{c^2}\left(\frac{1}{a}\frac{da}{dt}\right)^2-3\left(\frac{1}{a}\frac{da}{dt}\right)\,\frac{\partial}{\partial t}\left(\frac{\Phi}{c^2}\right)=4\pi G \delta\rho_{m},
\label{PotEq2}
\end{equation}
These equations do not capture all details of the $\Lambda$CDM model, but are sufficient for the following arguments.

The $\Lambda$CDM model explains a multitude of observed phenomena and, despite a couple of ``tensions'' in recent decades, nothing has yet falsified the model. But this author is struck by one feature of the model which seems a bit odd, and it is this oddness of the source term (\ref{SourceN}) which motivates construction of the $\Lambda$CDM-NG model.

%***************************
\subsection{The Source of the Gravitation Potential in the $\Lambda$CDM Model}
%**************************
Though not often emphasized, the equation for the potential in the $\Lambda$CDM model, Eq.(\ref{PotEq2}), is a modification of Newtonian gravitation, even in limits where Newtonian theory is expected to apply in general relativity. On the scale of galaxies and galaxy clusters, the terms containing $(1/a)(da/dt)=H\sim 2\times 10^{-18}\ s^{-1}$ in Eq.(\ref{PotEq2}) are negligible, and the term $\mathbf{\nabla}^2\Phi/a^2$ is actually the Newtonian Laplacian acting on $\Phi$ written in terms of comoving coordinates. So, over small regions, Eq.(\ref{PotEq2}) is the Newtonian Poisson equation, \emph{except that $\delta\rho_{m}$ appears as source term instead of the full density of condensed objects as in Newtonian theory}. The source term $\delta\rho_{m}$ has a number of curious properties which we list below:

\begin{itemize}
\item We are accustomed to thinking of the source term in the Newtonian field equation $\mathbf{\nabla}^2\Phi=4\pi G\rho_{\Phi}$ as the \textit{density of active gravitational mass}. But the space average of the density $\rho_{\Phi}=\delta\rho_{m}$ in Eq.(\ref{SourceN}) is zero. So the active gravitational mass density $\rho_{\Phi}(\mathbf{x},t)$ of the $\Lambda$CDM model is as much negative as it is positive. That is to say, there are regions where $\rho_{\Phi}(\mathbf{x},t)$ is negative and these regions have repulsive gravitational fields. In addition, the \emph{total} active gravitational mass is zero in this model!

\item If we move from cosmology to a different page of the general relativity textbook, we learn that the total mass of an isolated, gravitationally bound system (including the gravitational binding energy) may be determined using flux integral methods, if we know the weak asymptotic gravitational field far from the system \citep{Wheeler}. Such methods are used to establish the masses of black holes and neutron stars. But, if there exist a uniform negative gravitational mass density $-\bar{\rho}_{m}$ throughout space, as in Eq.(\ref{SourceN}), such methods will not work. The flux integral over a spherical surface, say, will necessarily go to zero as the radius tends to infinity, because the total active gravitational mass measured in this way is zero for the $\Lambda$CDM model.

\item Finally, it is odd that \textit{the source of the gravitational potential} at point $P$ should depend on masses at a great distance from $P$. This is indeed the case for the $\Lambda$CDM model because the source term at $P$, namely Eq.(\ref{SourceN}), depends on the average density $\bar{\rho}_{m}$ which in turn depends on masses at all distances from $P$. 

\end{itemize}

None of these comments rises to the status of a serious objection to the $\Lambda$CDM model. And, as a practical matter, the subtraction of $\bar{\rho}_{m}$ in the source term $\rho_{N}$ is of no concern for the $\Lambda$CDM model because $\bar{\rho}_{m}$ is very small (of the order of the critical density $\rho_{cr}=9.5\times 10^{-27} kg\,m^{-3}$) and is negligible compared to typical densities of stars, galaxies, and dark-matter clouds. But these thoughts do motivate one to ask: ``Is there a reasonable cosmological model which does not require a modification of Newtonian gravitation?''  The affermative answer to this question is the $\Lambda$CDM-NG model; a topic to which we now turn.

%***********************
\section{The Lambda-CDM-NG Model}
%***********************
The $\Lambda$CDM-NG model assumes a relativistic theory of gravitation should reduce to Newtonian theory in the limit where the latter is expected to apply. Indeed, Einstein made this very assumption when choosing the coupling constant between matter and gravitational field, so that his theory would reduce to Newtonian theory in the weak-field static limit \citep{Einstein1}. But, as we have seen, the $\Lambda$CDM model does not have the Newtonian limit, because its source for the gravitational potential is $\delta\rho_{m}$, and not the full mass density of condensed masses. The $\Lambda$CDM-NG model returns the source of gravitational potential to its full Newtonian value using the freedom to move mass between source terms established in Sec.2 (the NG in the moniker $\Lambda$CDM-NG remind us of the source term from \textbf{N}ewtonian \textbf{G}ravitation). According to Eq.(\ref{Constraint}), this step necessitates a decrease in the source term of Friedmann's equation.

The $\Lambda$CDM-NG model is determined by specifying the source terms $\rho_{F}(\tilde{t})$ and $\rho_{\Phi}(\tilde{\mathbf{x}},\tilde{t})$ for field equations (\ref{FriedGot}) and (\ref{PotEqGen}). If $\mathcal{F}(\tilde{t})$ is the fraction of matter which has condensed out of the near-uniform background density, then the average density of condensed objects is
\begin{equation}
\bar{\rho}_{\Phi}(\tilde{t})=\mathcal{F}(\tilde{t})\bar{\rho}_{m}(\tilde{t}),
\label{AveCon}
\end{equation}
and this is the \emph{space-averaged} source of gravitational potential in the $\Lambda$CDM-NG model.

Models of gravitational clustering, such as the collapse of a spherical slight overdensity \citep{Baumann}, suggest that the background density deviates very little from uniformity as structure formation progresses, and so, we expect the background density of matter to decrease with time as $[1-\mathcal{F}(\tilde{t})]\bar{\rho}_{m}(\tilde{t})$, and for this to be the contribution to $\rho_{F}$ from matter, 
\begin{equation}
\rho_{F}(\tilde{t})=\rho_{\Lambda}+[1-\mathcal{F}(\tilde{t})]\bar{\rho}_{m}(\tilde{t})+\rho_{r}(\tilde{t}), 
\label{SorF} 
\end{equation}
Then Eq.(\ref{Constraint}) determines the full source for the gravitational potential,
\begin{equation}
\rho_{\Phi}(\tilde{\mathbf{x}},\tilde{t})=\bar{\rho}_{\Phi}(\tilde{t})+[\rho_{m}(\tilde{\mathbf{x}},\tilde{t})-\bar{\rho}_{m}(\tilde{t})].
\label{SorN}
\end{equation}
Density $\rho_{\Phi}(\tilde{\mathbf{x}},\tilde{t})$ is the density of active gravitational mass, which we insist must be a positive quantity in the $\Lambda$CDM-NG model. For now we shall work with Eqs.(\ref{SorF}) and (\ref{SorN}), but $\rho_{\Phi}(\tilde{\mathbf{x}},\tilde{t})$ and $\mathcal{F}(\tilde{t})$ are not yet fully defined. In section 9 we give more precise definitions of these quantities in terms of the Fourier components of $\delta\rho_{m}$; definitions which garentee $\rho_{\Phi}(\tilde{\mathbf{x}},\tilde{t})\ge 0$. 

A word on notation. In the section after next, it is shown that the coordinates $\tilde{t}$ and $\tilde{x}^{i}$ we are using for the $\Lambda$CDM-NG model (those with a tilde ``$\ \tilde{}\ $'') are not the cosmic time $t$ and comoving coordinates $x^{i}$ used by astronomers to report their observations; the symbols $t$ and $x^{i}$ (or $\mathbf{x}$) are reserved for this purpose. In addition, we use the tilde to denote quantities that are not directly observed or measured. For example, the scale factor $a(t)$ in the $\Lambda$CDM model is directly observable, because $a=1/(1+z)$ and the redshift $z$ is observable. But in the $\Lambda$CDM-NG model, as we shall see, the scale factor $\tilde{a}$ is not directly observable, hence the notation.

Using standard formulas $\rho_{\Lambda}=$constant, $\rho_{r}(\tilde{t})=\rho_{r}(\tilde{t}_{0})/\tilde{a}^4$, and $\bar{\rho}_{m}(\tilde{t})=\bar{\rho}_{m}(\tilde{t}_{0})/\tilde{a}^3$ in Eqs.(\ref{SorF}) and (\ref{SorN}), and the results in Eqs.(\ref{FriedGot}) and (\ref{PotEqGen}), we obtain the basic equations of the $\Lambda$CDM-NG model:
\begin{equation}
ds^2=-\left(1+\frac{2\Phi}{c^2}\right)c^2d\tilde{t}^2+\tilde{a}^{2}(\tilde{t})\left(1-\frac{2\Phi}{c^2}\right)\delta_{ij}d\tilde{x}^{i} d\tilde{x}^{j},
\label{ScalarPe3}
\end{equation}
\begin{equation}
\tilde{\mathcal{H}}(\tilde{a})\equiv\left(\frac{1}{\tilde{a}}\frac{d\tilde{a}}{d\tilde{t}}\right)=H_{0}\sqrt{\Omega_{\Lambda}+\frac{[1-\mathcal{F}(\tilde{a})]\,\Omega_{m}}{\tilde{a}^3}+\frac{\Omega_{r}}{\tilde{a}^4}},
\label{ModFried2}
\end{equation}
\begin{equation}
\frac{1}{1+2\Phi/c^2}\left[\tilde{\mathcal{H}}-\frac{1}{1-2\Phi/c^2}\frac{\partial}{\partial\tilde{t}}\left(\frac{\Phi}{c^2}\right)\right]^2-\tilde{\mathcal{H}}^2+\frac{1}{(1-2\Phi/c^2)^2}\left[\frac{2}{3}\frac{\tilde{\mathbf{\nabla}}^2\Phi}{\tilde{a}^2}+\frac{\tilde{\nabla}\Phi\mathbf{\cdot}\tilde{\mathbf{\nabla}}\Phi}{c^2\tilde{a}^2 (1-2\Phi/c^2)}\right]=\frac{H^2_{0}\mathcal{F}(\tilde{t})\Omega_{m}}{\tilde{a}^3}+\frac{8\pi G}{3}\delta\rho_{m},
\label{PotEqGen2}
\end{equation}
where, as usual, $\Omega_{x}=\rho_{x}/\rho_{cr}$, $\rho_{cr}=3H_{0}/8\pi G$, and
\begin{equation}
H_{0}=\sqrt{8\pi G\rho_{cr}/3}
\label{HOh}
\end{equation}
is the \emph{observed Hubble constant}, and now $(\tilde{\mathbf{\nabla}}\Phi)_{i}=\partial\Phi/\partial \tilde{x}^{i}$, and $\tilde{\mathbf{\nabla}}^2\Phi=\delta^{ij}\partial^2\Phi/\partial \tilde{x}^{i}\partial \tilde{x}^{j}$.

The monotonically increasing condensed fraction $\mathcal{F}(\tilde{t})$ appearing in these equations is essentially zero in the very early universe, is expected to start growing when the radiation era transitions into the matter era, grows in earnest after decoupling, and finishes at some final value $\mathcal{F}_{0}=\mathcal{F}(\tilde{t}_{0})$ at the present time $\tilde{t}_{0}$. It is often more convenient, as in Eq.(\ref{ModFried2}), to write $\mathcal{F}[\tilde{a}(\tilde{t})]$ as a function of the scale factor $\tilde{a}(\tilde{t})$, in which case $\mathcal{F}_{0}=\mathcal{F}(\tilde{a}_{0})$.

%*************************
\section{The Gravitational Potential of the Universe}
%*************************
Solution of the rather cumbersome equation (\ref{PotEqGen2}) for potential $\Phi$ is simplified by initially ignoring the inhomogeneous part $\delta\rho_{m}$ of the source term, and writing this equation for the position-independent potential $\bar{\Phi}(\tilde{t})$ produced by the average density $\bar{\rho}_{\Phi}=\mathcal{F}(\tilde{t})\bar{\rho}_{m}(\tilde{t}_{0})/\tilde{a}^3(\tilde{t})$ alone, namely
\begin{equation}
\frac{1}{1+2\bar{\Phi}/c^2}\left[\tilde{\mathcal{H}}-\frac{1}{1-2\bar{\Phi}/c^2}\frac{d}{d\tilde{t}}\left(\frac{\bar{\Phi}}{c^2}\right)\right]^2=\tilde{\mathcal{H}}^2+\frac{H^2_{0}\mathcal{F}(\tilde{t})\Omega_{m}}{\tilde{a}^3}.
\label{Interum}
\end{equation}
Solving this equation for $d(\bar{\Phi}/c^2)/d\tilde{t}$ and expanding $\tilde{\mathcal{H}}$ using Eq.(\ref{ModFried2}), we have
\begin{equation}
\frac{d}{d\tilde{t}}\left(\frac{\bar{\Phi}}{c^2}\right)=-H_{0}\left(1-\frac{2\bar{\Phi}}{c^2}\right)
\left(J_{1}-J_{2}\right),
\label{PotDiriv}
\end{equation}
where
\begin{equation}
J_{1}=\sqrt{\left(1+\frac{2\bar{\Phi}}{c^2}\right)\left(\Omega_{\Lambda}+\frac{\Omega_{m}}{\tilde{a}^3}+\frac{\Omega_{r}}{\tilde{a}^4}\right)},
\label{Part1}
\end{equation}
and
\begin{equation}
J_{2}=\sqrt{\Omega_{\Lambda}+\frac{[1-\mathcal{F}(\tilde{a})]\Omega_{m}}{\tilde{a}^3}+\frac{\Omega_{r}}{\tilde{a}^4}}.
\label{Part2}
\end{equation}
This is our working equation for $\bar{\Phi}/c^2$; a quantity we call \emph{the gravitational potential of the universe}.

It is then straightforward, using Eq.(\ref{PotEqGen2}), to write an equation for the first order perturbation $\delta\Phi$ to potential $\bar{\Phi}$ produced by the inhomogeneous part $\delta\rho_{m}$ of the source density in Eq.(\ref{PotEqGen2}); an equation important for structure formation in the $\Lambda$CDM-NG model. However, the principle purpose of this paper is to show how the potential $\bar{\Phi}(\tilde{t})$ could resolve the Hubble tension, and we shall not consider $\delta\Phi$ further at this time.

%******************************
\section{The Observed Scale Factor}
%******************************
In a simple line element of the form
\begin{equation}
ds^2=-c^2dt^2+a^2(t)\delta_{ij}dx^{i}dx^{j},
\label{Simp}
\end{equation}
$t$ is the proper time reading on synchronized comoving clocks known as \emph{cosmic time}, and the comoving space coordinates $x^{i}$ are proper distances measured in the coordinate directions at the present time $t_{0}$. Such interpretations are critical for relating the formalism to observations.

When the gravitational potential $\Phi$ is added to Eq.(\ref{Simp}) to obtain a line element of the form
\begin{equation}
ds^2=-\left(1+\frac{2\Phi}{c^2}\right)c^2d\tilde{t}^2+\tilde{a}^{2}(\tilde{t})\left(1-\frac{2\Phi}{c^2}\right)\delta_{ij}d\tilde{x}^{i} d\tilde{x}^{j},
\label{ScalarPe4}
\end{equation}
the time coordinate $\tilde{t}$ is no longer the cosmic time $t$ used by astronomers to report their results, and the space coordinates $\tilde{x}_{i}$ are not, at the present time, proper distances in the coordinate directions. In the $\Lambda$CDM model, where $\Phi/c^2<<1$, this distinction is unimportant, and $\tilde{t}$ and $\tilde{x}^{i}$ can be treated as if these were the physical variables $t$ and $x^{i}$. But in the $\Lambda$CDM-NG model, where $\Phi/c^2$ can be larger, such an approximation may be inappropriate.

For line element (\ref{ScalarPe4}), the increment of proper time on a comoving clock (where $d\tilde{x}^{i}=0$) is 
\begin{equation}
dt=\sqrt{1+2\bar{\Phi}(\tilde{t})/c^2}d\tilde{t},
\label{PropInc}
\end{equation}
and
\begin{equation}
t(\tilde{t})=\int_{0}^{\tilde{t}}\sqrt{1+2\bar{\Phi}(\tilde{t^{'}})/c^2}\,d\tilde{t^{'}}.
\label{PropTime}
\end{equation}
Note that $t(\tilde{t})$ is a strictly increasing function of $\tilde{t}$ which has inverse $\tilde{t}(t)$. 

Similarly, the increments of proper distance in the coordinate directions, at the present time $\tilde{t}_{0}$, when $\tilde{a}(\tilde{t}_{0})=1$, are
\begin{equation}
dx^{i}=\sqrt{1-2\bar{\Phi}(\tilde{t}_{0})/c^2}d\tilde{x}^{i},
\label{PropDist}
\end{equation}
and
\begin{equation}
\tilde{x}^{i}=\frac{x^{i}}{\sqrt{1-2\bar{\Phi}(\tilde{t}_{0})/c^2}}.
\label{PhysDist}
\end{equation}
Equations (\ref{PropTime}) and (\ref{PhysDist}) constitute a coordinate transformation to physical coordinates $t$ and $x^{i}$, which recasts line element (\ref{ScalarPe4}) into the form
\begin{equation}
ds^2=-c^2dt^2+a_{obs}^2(t)\delta_{ij}dx^{i}dx^{j},
\label{ObsMet}
\end{equation}
where 
\begin{equation}
a_{obs}(t)=\tilde{a}(\tilde{t}(t))\sqrt{\frac{1-2\bar{\Phi}(\tilde{t}(t))/c^2}{1-2\bar{\Phi}(\tilde{t}(t_{0}))/c^2}},
\label{ObsScale}
\end{equation}
is the \emph{observed scale factor}, because the coordinates $t$ and $x^{i}$ used in line element (\ref{ObsMet}) are now the measured spacetime coordinates used by astronomers. Without this connection to measured quantities, the equations of cosmology would be empty formalism.

From the observed scale factor, one calculates the \emph{observed Hubble parameter} as
\begin{equation}
H^{obs}(t)=\frac{1}{a_{obs}}\frac{da_{obs}}{dt},
\label{HubblePar}
\end{equation}
using cosmic time $t$; and the \emph{observed deceleration parameter} as
\begin{equation}
q^{obs}(t)=-\frac{1}{[H^{obs}(t)]^2}\left(\frac{1}{a_{obs}(t)}\frac{d^2a_{obs}}{dt^2}\right)=-\left[1+\frac{1}{(H^{obs})^2}\frac{dH^{obs}}{dt}\right],
\label{AccelParam}
\end{equation}

%*****************************
\section{The ``Friedmann Equation''}
%*****************************

In the $\Lambda$CDM-NG model the ``Friedmann equation,'' reads
\begin{equation}
\frac{1}{\tilde{a}}\frac{d\tilde{a}}{d\tilde{t}}=H_{0}\sqrt{\Omega_{\Lambda}+[1-\mathcal{F}(\tilde{a})]\frac{\Omega_{m}}{\tilde{a}^3}+\frac{\Omega_{r}}{\tilde{a}^4}}.
\label{NGFried}
\end{equation}
where the quantity on the right is less than that in the corresponding $\Lambda$CDM model, Eq.(\ref{Friedmann2}), because $[1-\mathcal{F}(\tilde{a})]$ is less than one. With $\Omega_{\Lambda}+\Omega_{m}+\Omega_{r}=1$ and $\tilde{a}(\tilde{t}_{0})=1$ at the present time $\tilde{t}_{0}$, Eq.(\ref{NGFried}) would seem to be inconsistent, because the right side of this equation is less than the Hubble constant. But this is not an inconsistency, because, as we have seen, $\tilde{a}(\tilde{t})$ is not the observed scale factor.

From the results of the preceding section we conclude that the role of Eq.(\ref{NGFried}) in the $\Lambda$CDM-NG is different than in the $\Lambda$CDM model, in that it does not directly determine the observed Hubble parameter. We shall refer to $\tilde{a}(\tilde{t})$ as the \emph{prior scale factor}, and we call 
\begin{equation}
\tilde{\mathcal{H}}(\tilde{a})=H_{0}\sqrt{\Omega_{\Lambda}+[1-\mathcal{F}(\tilde{a})]\frac{\Omega_{m}}{\tilde{a}^3}+\frac{\Omega_{r}}{\tilde{a}^4}}
\label{PriorH}
\end{equation}
 the \emph{prior Hubble parameter} to distinguish this from the \emph{observed Hubble parameter} $H^{obs}(t)$ which we discuss next.

%*******************************
\section{The Observed Hubble Parameter and Hubble Constant}
%******************************
We are now in a position to evaluate the observed Hubble parameter. Taking the derivative of $a_{obs}(t)$, Eq.(\ref{ObsScale}), with respect to cosmic time $t$, and dividing by $a_{obs}(t)$, we have the observed Hubble parameter of Eq.(\ref{HubblePar}),
\begin{equation}
H^{obs}(t)=\left[\frac{1}{\tilde{a}}\frac{d\tilde{a}}{d\tilde{t}}-\frac{1}{1-2\bar{\Phi}/c^2}\frac{d}{d\tilde{t}}\left(\frac{\bar{\Phi}}{c^2}\right)\right]\frac{d\tilde{t}}{dt}.
\label{ObsHubble}
\end{equation}
Then using Eqs.(\ref{ModFried2}), (\ref{PropInc}), and (\ref{PotDiriv}), Eq.(\ref{ObsHubble}) becomes
\begin{equation}
H^{obs}(t)=H_{0}\sqrt{\Omega_{\Lambda}+\frac{\Omega_{m}}{\tilde{a}^3}+\frac{\Omega_{r}}{\tilde{a}^4}}.
\label{HubObs}
\end{equation}
It appears that the density we had subtracted from the source term in Friedman's equation (\ref{ModFried2}) is restored to the Friedmann equation (\ref{HubObs}) when working with observed quantities. But this is not exactly correct because the $\tilde{a}$ in Eq.(\ref{HubObs}) is not the observed scale factor. Using Eq.(\ref{ObsScale}) to write Eq.(\ref{HubObs}) in terms of the observed scale factor, we obtain
\begin{equation}
H^{obs}(t)=H_{0}\sqrt{\Omega_{\Lambda}+\Omega_{m}\left(\frac{X}{a_{obs}}\right)^3+\Omega_{r}\left(\frac{X}{a_{obs}}\right)^4},
\label{HubObs2}
\end{equation}
where
\begin{equation}
X(t)=\sqrt{\frac{1-2\bar{\Phi}(t)/c^2}{1-2\bar{\Phi}(t_{0})/c^2}}.
\label{PhiThing}
\end{equation}
Note that, at the present time when $a_{obs}=1$ and $X=1$, Eq.(\ref{HubObs2}) gives
\begin{equation}
\Omega_{\Lambda}+\Omega_{m}+\Omega_{r}=1,
\label{StanRe2}
\end{equation}
as in the $\Lambda$CDM model. Therefore, the observed Hubble parameter at the present time in the $\Lambda$CDM-NG model (the ``Hubble constant'' in this model), is 
\begin{equation}
H^{obs}(t_{0})=H_{0}=\sqrt{\frac{8\pi G\rho_{cr}}{3}},
\label{HubConNG}
\end{equation}
as in the $\Lambda$CDM model. \emph{But, at other times, the observed Hubble parameter $H^{obs}(t)$ differs from that of the $\Lambda$CDM model.} Note also that this result is exact for line element (\ref{ScalarPe2}) when fluctuations of the density are ignored. Finally, we see that the $\Lambda$CDM-NG model differs from the $\Lambda$CDM model only by the appearance of the potential factor (\ref{PhiThing}) in the Friedmann equation (\ref{HubObs2}); a factor which differs from unity in the $\Lambda$CDM-NG model.

%*************************
\section{Two Parameters Defined}
%*************************
In this section we define $\rho_{\Phi}(\mathbf{x},t)$ and $\mathcal{F}(t)$ in terms of the Fourier components of $\delta\rho_{m}(\mathbf{x}, t)$. This brings the definitions of these quantities closer to certain known quantities such as the power spectrum of density fluctuations.

Consider the simple density $\rho_{m}(x)$ depicted in Figure 1. It is a harmonic function of $x$ of amplitude $r$ raised above the zero level by amount $\bar{\rho}_{m}$. In the $\Lambda$CDM model, this density would be split into an average density $\bar{\rho}_{m}$ and a harmonic ``ripple'' $\delta\rho_{m}=r\,cos(kx+\phi)$ of average value zero. The average density $\bar{\rho}_{m}$ would contribute to the source term in Friedmann's equation, and the ripple would be the source of gravitational potential $\Phi$ in this model.

In the $\Lambda$CDM-NG model, on the other hand, the source of the gravitational potential (the \emph{density of active gravitational mass}) is required to be positive. Accordingly, the density $\rho_{m}(x)$ in the figure is split by the grid line at the bottom of the ripple. The density above this line $\rho_{\Phi}(x)$ is the source of gravitational potential in the $\Lambda$CDM-NG model, and the position-independent density below this line, $(1-\mathcal{F}\,)\bar{\rho}_{m}$, contributes to the source term in Friedmann's equation. Notice that the source of potential in the $\Lambda$CDM-NG model is the source $\delta\rho_{m}$ in the $\Lambda$CDM model plus the amplitude $r$ of the ripple,
\begin{eqnarray}
\rho_{\Phi}(x)&=&r+r\,cos(kx+\phi), \nonumber \\
&=&r+\delta\rho_{m}.
\label{AddAmp}
\end{eqnarray}
This is more than a simplified pedagogical example, because each Fourier component of a realistic density has a profile like that in Figure 1.

We expand a general source of gravitational potential $\delta\rho_{m}(\mathbf{x},t)$ in the $\Lambda$CDM model as the Fourier integral
\begin{equation}
\delta\rho_{m}(\mathbf{x},t)=\frac{1}{(2\pi)^3}\int\delta\rho_{m}(\mathbf{k},t)e^{i\mathbf{k}\cdot\mathbf{x}}d^3k,
\label{FourTrans}
\end{equation}
with inverse transform
\begin{equation}
\delta\rho_{m}(\mathbf{k},t)=\int\delta\rho_{m}(\mathbf{x},t)e^{-i\mathbf{k}\cdot\mathbf{x}}d^3x.
\label{InvFourTran}
\end{equation}
Because $\delta\rho_{m}(\mathbf{x},t)$ is real, Eq.(\ref{FourTrans}) can be written as
\begin{equation}
\delta\rho_{m}(\mathbf{x},t)=\int\, r(\mathbf{k},t)\,cos[\mathbf{k}\cdot\mathbf{x}+\phi(\mathbf{k},t)]d^3k,
\label{Ripple}
\end{equation}
where we have written $\delta\rho_{m}(\mathbf{k},t)=|\delta\rho_{m}(\mathbf{k},t)|e^{i\phi(\mathbf{k},t)}$ in terms of modulus and phase, and have set
\begin{equation}
r(\mathbf{k},t)\equiv\frac{|\delta\rho_{m}(\mathbf{k},t)|}{(2\pi)^3}\ge 0.
\label{AmpDef}
\end{equation}
Equation (\ref{Ripple}) is a sum of functions of the form $r(\mathbf{k},t)\,cos[\mathbf{k}\cdot\mathbf{x}+\phi(\mathbf{k},t)]$, each of which has average value zero, and, if the $x$ axis is chosen in the direction of $\mathbf{k}$, has the same form as in our simple example above. By analogy with that example, we conclude that the source term for the potential in the $\Lambda$CDM-NG model is obtained by adding the amplitude $r(\mathbf{k},t)$ to the integrand in Eq.(\ref{Ripple}),
\begin{eqnarray}
\rho_{\Phi}(\mathbf{x},t)&=&\int\,r(\mathbf{k},t)(1+ cos[\mathbf{k}\cdot\mathbf{x}+\phi(\mathbf{k},t)])d^3k \nonumber \\
&=&\int\,r(\mathbf{k},t)d^3k+\delta\rho_{m}(\mathbf{x},t).
\label{NGSource}
\end{eqnarray}
A space average of this result, with $<\delta\rho_{m}>=0$, gives
\begin{equation}
\bar{\rho}_{\Phi}(t)=\int\,r(\mathbf{k},t)d^3k,
\label{AveRhoN}
\end{equation}
and using $\bar{\rho}_{\Phi}=\mathcal{F}(\tilde{t})\bar{\rho}_{m}$, we have
\begin{equation}
\mathcal{F}(t)=\frac{\int\,r(\mathbf{k},t)d^3k}{\bar{\rho}_{m}(t)}=\frac{1}{(2\pi)^3\bar{\rho}_{m}(t)}\int\,|\delta\rho_{m}(\mathbf{k},t)|d^3k.       
\label{ConFrac}
\end{equation}
Eqs.(\ref{NGSource}) and (\ref{ConFrac}) are our definitions of $\rho_{\Phi}(\mathbf{x},t)$ and $\mathcal{F}(t)$, and Eqs.(\ref{AmpDef}) and (\ref{AveRhoN}) ensure that $\rho_{\Phi}\ge 0$.

\begin{figure}
\begin{center}
\includegraphics[width=0.60\textwidth]{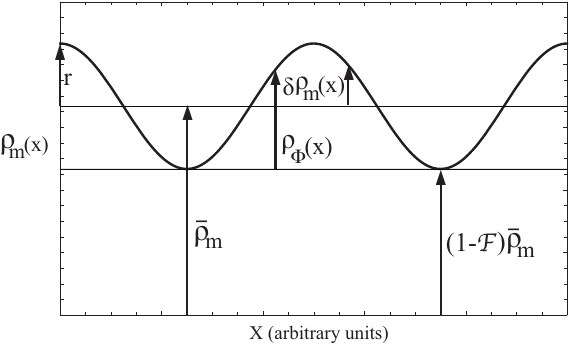}
\end{center}
\begin{quote}
\caption{\label{Fig.1} Schematic diagram of a density $\rho_{m}(x)$ consisting of a harmonic ripple on top of a constant density. The relationships of densities $\bar{\rho}_{m}$, $\delta\rho(x)$, $\rho_{\Phi}(x)$, and $(1-\mathcal{F}\,)\bar{\rho}_{m}$ are shown for this total density.}
\end{quote}
\end{figure}

Although the condensation fraction in Eq.(\ref{ConFrac}) is clearly related to the power spectrum of density fluctuations $\mathcal{P}(k)$, it is not derivable from the latter, and, at the present time, the form of $\mathcal{F}(t)$ for our universe is not known with any confidence. However, as shown in Sec.13, there is quite a variety of condensation histories for the accepted values of density parameters $\Omega_{\Lambda}, \Omega_{m}$, and $\Omega_{r}$, which give the observed values for $H_{0}$ and $q_{0}$ and, at the same time, resolve the Hubble tension within the $\Lambda$CDM-NG model.

%*************************
\section{The Early Universe}
%*************************
Here the term ``early universe'' is taken to mean the period from the big bang ($t=0$) up to and including the time of photon decoupling or last scattering ($t_{ls}\approx 3.7\times 10^{5}$ yr, $z_{ls}=1090$, $a_{ls}=9.2\times 10^{-4}$). Our conclusion will be: There is essentially no difference between the $\Lambda$CDM and $\Lambda$CDM-NG models during this period.

On comparing $\bar{\rho}_{\Phi}$, Eq.(\ref{AveRhoN}), with $\delta\rho_{m}$, Eq.(\ref{Ripple}), we see that, although $\bar{\rho}_{\Phi}$ is greater than $\delta\rho_{m}$ (because $1\ge cos\theta$), the two quantities are of the same order of magnitude. Therefore, the density contrast $\delta=\delta\rho_{m}/\bar{\rho}_{m}$ in the $\Lambda$CDM model is of the same order of magnitude as the condensed fraction $\mathcal{F}=\bar{\rho}_{N}/\bar{\rho}_{m}$ in the $\Lambda$CDM-NG model. Estimates of $\delta$ at last scattering are of order $\delta\sim 10^{-3}$ or smaller \citep{Weinberg3}, and a condensed fraction of this order is negligible in the prior Friedmann equation (\ref{NGFried}); which puts this equation into the same form as the Friedmann equation of the $\Lambda$CDM model (\ref{Friedmann2}). 

With sources, $\delta\rho_{m}$ and $\rho_{\Phi}$, for the gravitational potentials in the two models being of the same order of magnitude, the two potentials are also of the same magnitude. From \emph{Sachs-Wolfe} theory, we know the dimensionless potential $\Phi/c^2$ is of the same order of magnitude as the temperature fluctuations of the CMB radiation ($\Phi/c^2\sim\Delta T/\bar{T}\sim 10^{-5}$) \citep{Sachs}. Hence, the potential $\bar{\Phi}(\tilde{t}(t))/c^2$ at early time $t$ in the observed scale factor (\ref{ObsScale}) is negligible, and the observed scale factor $a_{obs}(t)$ differs from $\tilde{a}(t)$ only by the constant factor $1/\sqrt{1-2\bar{\Phi}(\tilde{t}(t_{0}))/c^2}$, which has no effect on the Hubble parameter. Consequently, the Hubble parameters in the two models are equal [$H^{obs}(t)=H(t)$] and, in particular, the Hubble parameters at last scattering are equal [$H^{obs}(a_{ls})=H(a_{ls})$], a fact used in the following section. The small value of the potential $\bar{\Phi}/c^2$ at early time also implies, through Eq.(\ref{PropTime}), a negligible difference between time scales $\tilde{t}$ and $t$ in the early universe.

In short, there is no significant difference between the $\Lambda$CDM and $\Lambda$CDM-NG models at and before last scattering. Therefore, the various phenomena occurring before $t_{ls}$, such as the creation of light elements, which are cited as evidence for the $\Lambda$CDM model, are also consistent with the $\Lambda$CDM-NG model. 

%*******************************
\section{The Hubble Tension}
%******************************
As noted in the previous section, the $\Lambda$CDM and $\Lambda$CDM-NG models are essentially the same in the early universe, with the same value for the Hubble parameter at last scattering:
\begin{equation}
H(a_{ls})=(15.24\pm 0.12)\times 10^5\ \mathrm{km\,s^{-1}\,Mpc^{-1}}.
\label{HDec}
\end{equation}
But, after this time, the models differ. In the $\Lambda$CDM model the Hubble parameter at last scattering is extrapolated to the Hubble parameter at the present time $H_{0}$ using the Friedmann equation for this model,
\begin{equation}
H_{0}=\frac{H(a_{ls})}{\sqrt{\Omega_{\Lambda}+\Omega_{m}/a^3_{ls}+\Omega_{r}/a^{4}_{ls}}}.
\label{HExtrap}
\end{equation}
The value of the Hubble constant obtained in this way, $H_{0}=67.4\pm 0.5\ \mathrm{km\,s^{-1}\,Mpc^{-1}}$, and reported by the Planck Collaboration \citep{Planck}, is in tension with the late-universe, distance-ladder result of about $H_{0}= 73.04\pm 1.04\ \mathrm{km\,s^{-1}\,Mpc^{-1}}$ \citep{Riess}.

But the $\Lambda$CDM-NG model has a different extrapolation formula. From Eq.(\ref{HubObs2}) we have
\begin{equation}
H_{0}=\frac{H^{obs}(a^{obs}_{ls})}{\sqrt{\Omega_{\Lambda}+\Omega_{m}\left(X_{ls}/a^{obs}_{ls}\right)^3+\Omega_{r}\left(X_{ls}/a^{obs}_{ls}\right)^4}},
\label{NGExtrap}
\end{equation}
where $X_{ls}$ is given by Eq.(\ref{PhiThing}) with $a_{obs}=a^{obs}_{ls}$. The potential $\bar{\Phi}(a^{obs}_{ls})/c^2$ in  $X_{ls}$ is always small compared to unity (see previous section). Therefore, to good approximation,
\begin{equation}
X_{ls}\approx\frac{1}{\sqrt{1-2\bar{\Phi}_{0}/c^2}},
\label{XApprox}
\end{equation}
and $X_{ls}$ depends only on the potential $\bar{\Phi}_{0}/c^2$ at the present time.

\begin{figure}
\begin{center}
\includegraphics[width=0.60\textwidth]{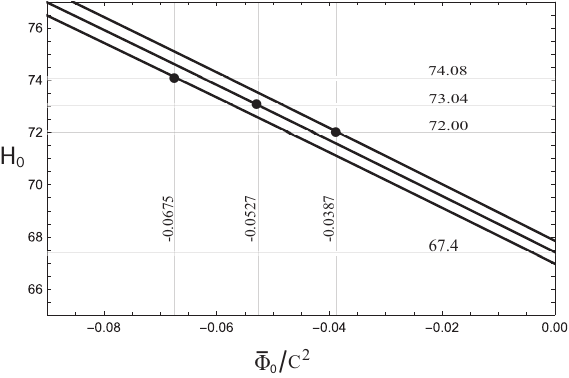}
\end{center}
\begin{quote}
\caption{\label{Fig.2} The Hubble constant of the $\Lambda$CDM-NG model is plotted as a function of the dimensionless gravitational potential at the present time $\bar{\Phi}_{0}/c^2$. The three curves, which begin at $H_{0}=67.9,\ 67.4,$ and $\ 66.9\ \mathrm{km\,s^{-1}\,Mpc^{-1}}$ at $\bar{\Phi}_{0}/c^2=0$, show how the $\Lambda$CDM-NG Hubble constant increases as the strength of the potential $\bar{\Phi}_{0}/c^2$ increases. The three bold dots on the trend lines mark the range of values of potential $\bar{\Phi}_{0}/c^2$ for which the $\Lambda$CDM-NG agrees with the late-universe, distance-ladder measurements of the Hubble constant.}
\end{quote}
\end{figure}

The Hubble constant in the $\Lambda$CDM-NG model (\ref{NGExtrap}) is plotted in Figure 2 as a function of $\bar{\Phi}_{0}/c^2$. At $\bar{\Phi}_{0}/c^2=0$, Eq.(\ref{NGExtrap}) reduces to Eq.(\ref{HExtrap}) of the $\Lambda$CDM model, and the three curves begin at the values $H_{0}=67.9,\ 67.4,\ 66.9\ \mathrm{km\,s^{-1}\,Mpc^{-1}}$ describing the uncertainty of $H_{0}$ in the $\Lambda$CDM model. As $\bar{\Phi}_{0}/c^2$ increases negatively to the left, the $H_{0}$ of the $\Lambda$CDM-NG model trends upward. The three horizontal grid lines at $H_{0}=72.00,\ 73.04,\ 74.08\ \mathrm{km\,s^{-1}\,Mpc^{-1}}$ bracket the range of values obtained in the late-universe measurements. The three bold dots on the curves mark the range of  $\bar{\Phi}_{0}/c^2$ for which there is no Hubble tension in the $\Lambda$CDM-NG model. Hence, the $\Lambda$CDM-NG model resolves the Hubble tension if $\bar{\Phi}_{0}/c^2$ falls in the range $-0.0675< \bar{\Phi}_{0}/c^2< -0.0387$. As we shall see in the section after next, a gravitational potential of this magnitude does not seem unreasonable for the actual universe.

%*******************************
\section{The Deceleration Parameter}
%******************************

To be considered seriously, a cosmological model must have a nearly flat spatial section, have a Hubble constant in the observed range, and a deceleration parameter consistent with the accelerated expansion of the universe. The $\Lambda$CDM-NG model begins with flat spatial geometry, and Eqs.(\ref{HubObs2}) automatically give the observed Hubble constant when $\Omega_{\Lambda}+\Omega_{m}+\Omega_{r}=1$. So it remains to show that the deceleration parameter in the $\Lambda$CDM-NG model is consistent with observations of $q_{0}$. 

From Eqs.(\ref{HubObs2}) and (\ref{PhiThing}) we can calculate the deceleration parameter in the form
\begin{equation}
q^{obs}=-\left[1+\frac{1}{(H^{obs})^2}\frac{dH^{obs}}{dt}\right],
\label{DecelP}
\end{equation}
with previous equations required in the calculation presenting themselves as the calculation proceeds. The result is
\begin{equation}
q^{obs}=-\left[\frac{K_{1}^2-Q(K_{2}/K_{1})}{K_{1}^2}\right],
\label{Decelq}
\end{equation}
where
\begin{equation}
Q=\frac{1}{\sqrt{1+2\bar{\Phi}/c^2}}\left[\frac{3}{2}\Omega_{m}\left(\frac{X}{a_{obs}}\right)^3+2\Omega_{r}\left(\frac{X}{a_{obs}}\right)^4\right],
\label{QPart}
\end{equation}
\begin{equation}
K_{1}=\sqrt{\Omega_{\Lambda}+\Omega_{m}\left(\frac{X}{a_{obs}}\right)^3+\Omega_{r}\left(\frac{X}{a_{obs}}\right)^4}.
\label{hAbreev2}
\end{equation}
\begin{equation}
K_{2}=\sqrt{\Omega_{\Lambda}+(1-\mathcal{F})\Omega_{m}\left(\frac{X}{a_{obs}}\right)^3+\Omega_{r}\left(\frac{X}{a_{obs}}\right)^4},
\label{hAbreev}
\end{equation}
As the condensed fraction $\mathcal{F}(\tilde{a})$ goes to zero (and $\bar{\Phi}$ vanishes as a result), deceleration parameter (\ref{Decelq}) becomes the deceleration parameter
\begin{equation}
q^{\Lambda CDM}=-\left[\frac{\Omega_{\Lambda}-(1/2)\Omega_{m}/a^3_{obs}-\Omega_{r}/a^4_{obs}}{\Omega_{\Lambda}+\Omega_{m}/a^3_{obs}+\Omega_{r}/a^4_{obs}}\right]
\label{lamCDMq}
\end{equation}
of the $\Lambda$CDM model. In fact, all results of the $\Lambda$CDM-NG model reduce to those of the $\Lambda$CDM model as $\mathcal{F}(\tilde{a})$ tends to zero.

At the present time, when $X=a_{obs}=1$, the deceleration parameter (\ref{Decelq}) becomes the ``deceleration constant''
\begin{equation}
q^{obs}_{0}=\frac{1}{2}\Omega_{m}+\Omega_{r}-\Omega_{\Lambda}+\left(\frac{3}{2}\Omega_{m}+2\Omega_{r}\right)\left(\sqrt{\frac{1-\mathcal{F}_{0}\Omega_{m}}{1+2\bar{\Phi}_{0}/c^2}}-1\right).
\label{DecelConst}
\end{equation}
Equation (\ref{DecelConst}), offers the opportunity to distinguish the $\Lambda$CDM-NG model from the $\Lambda$CDM model. The $q^{\Lambda CDM}_{0}=(1/2)\Omega_{m}+\Omega_{r}-\Omega_{\Lambda}$ deceleration constant depends on $\Omega_{m}$, $\Omega_{r}$, and $\Omega_{\Lambda}$ only. But if an independent measurement of $q_{0}$ is made with result differing from $q^{\Lambda CDM}_{0}$ by an amount consistent with Eq.(\ref{DecelConst}), one would have evidence supporting the $\Lambda$CDM-NG model. Example condensation histories $\mathcal{F}(\tilde{a})$ in the following section, having potentials $\bar{\Phi}_{0}/c^2$ sufficient to explain the Hubble tension, suggest the terms in parentheses in (\ref{DecelConst}) differs from $q^{\Lambda CDM}_{0}$ by one or two percent.

%*******************************
\section{Condensation Histories}
%******************************

We shall show, in this section, that the observed parameters $H_{0}$ and $q_{0}$ can be realized in the $\Lambda$CDM-NG model at the same time it resolves the Hubble tension, and that it does so for a variety of condensation histories $\mathcal{F}(\tilde{a})$, without the need for any fine tuning of parameters. Ideally, the other parameters in the $\Lambda$CDM-NG model, namely $\Omega_{m}$, $\Omega_{\Lambda}$, $\Omega_{r}$, should be obtained from a best fit of the $\Lambda$CDM-NG model to observational data, but we shall take these values from the Planck Collaboration results \citep{Planck}, because these values are constrained by a variety of independent observations  \citep{Perlmutter,Riess,Burles,Cooke,Abbott,Mantz}, and cannot deviate much from the Planck values.

Figure 3 shows three rather different condensation histories, all of which are deived from parameters $\Omega_{\Lambda}=0.684$, $\Omega_{m}=0.316$, $\Omega_{r}=7.7\times 10^{-5}$, and $H(a_{ls})$ from Eq.(\ref{HDec}). Curve (1) has $\bar{\Phi}_{0}/c^2=-0.0456$, $\mathcal{F}_{0}=0.35$, $H^{obs}_{0}=72.3\mathrm{\ km\,s^{-1}\,Mpc^{-1}}$, and $q^{obs}_{0}=-0.531$; Curve (2) has $\bar{\Phi}_{0}/c^2=-0.0400$, $\mathcal{F}_{0}=0.40$, $H^{obs}_{0}=71.7\mathrm{\ km\,s^{-1}\,Mpc^{-1}}$, and $q^{obs}_{0}=-0.538$; and Curve (3) has $\bar{\Phi}_{0}/c^2=-0.0418$, $\mathcal{F}_{0}=0.2$, $H^{obs}_{0}=71.9\mathrm{\ km\,s^{-1}\,Mpc^{-1}}$, and $q^{obs}_{0}=-0.521$. Thus all of the curves give deceleration parameters in the range $q_{0}=0.526\pm 0.013$, and all have mean gravitational potentials in the range $-0.0675< \bar{\Phi}_{0}/c^2< -0.0387$ which give Hubble constants for which there is no Hubble tension in the $\Lambda$CDM-NG model.

We conclude there is a broad range of condensation histories which resolve the Hubble tension in the $\Lambda$CDM-NG model, and no fine tuning of $\mathcal{F}(\tilde{a})$ is required to secure this result.

\begin{figure}
\begin{center}
\includegraphics[width=0.60\textwidth]{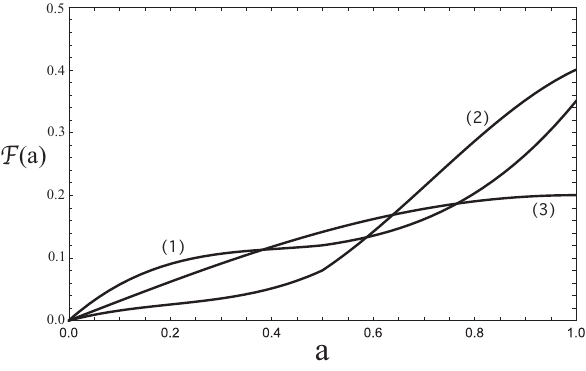}
\end{center}
\begin{quote}
\caption{\label{Fig.3} Three different condensation fractions $\mathcal{F}(\tilde{a})$ are plotted to illustrate the variety of condensation histories which resolve the Hubble tension in the $\Lambda$CDM-NG model, and have Hubble constants and deceleration constants in the observed range. One concludes that no fine tuning of the condensation fraction $\mathcal{F}(\tilde{a})$ is required to obtain the observed cosmological parameters in the $\Lambda$CDM-NG model.}

\end{quote}
\end{figure}

%*******************************
\section{Physical Interpretation of the $\Lambda$CDM-NG Model}
%******************************

When a clock is placed near a large mass, whose gravitational potential at the clock is $\Phi$, the time interval $d\tilde{t}$ the clock would have registered in the absence of the mass is shortened ($\Phi$ is negative) to the value
\begin{equation}
dt=\sqrt{1+\frac{2\Phi}{c^2}}d\tilde{t}.
\label{TimeDi}
\end{equation}
This is the well understood, and well tested, gravitational time dilation effect.

When the average density of condensed masses $\bar{\rho}_{\Phi}$ is included in the source term for the gravitational potential, it creates gravitational potential $\bar{\Phi}$, and the time $d\tilde{t}$ that would have evolved on the standard comoving clocks used by astronomers is shortened to the value
\begin{equation}
dt=\sqrt{1+\frac{2\bar{\Phi}}{c^2}}d\tilde{t}.
\label{TimeDiNG}
\end{equation}
This is Eq.(\ref{PropInc}) of Sec.6. 

Proper distance is also altered by gravitational time dilation, because proper distance is defined in terms of light travel time (local radar distance) as measured by a clock experiencing gravitational time dilation. As shown in Sec.6, this is the source of the factor
\begin{equation}
X(t)=\sqrt{\frac{1-2\bar{\Phi}(\tilde{t}(t))/c^2}{1-2\bar{\Phi}(\tilde{t}(t_{0}))/c^2}}
\label{XFactor}
\end{equation}
which multiplies the prior scale factor $\tilde{a}(\tilde{t}(t))$ to give the observed scale factor
\begin{equation}
a_{obs}(t)=\tilde{a}[\tilde{t}(t)]X[\tilde{t}(t)].
\label{ObsSF}
\end{equation}
Integration of Eq.(\ref{TimeDiNG}) determines the cosmic time
\begin{equation}
t(\tilde{t})=\int_{0}^{\tilde{t}}\sqrt{1+\frac{2\bar{\Phi}(\tilde{t}^{'})}{c^2}}d\tilde{t}^{'}
\label{PhysTime3}
\end{equation}
and its inverse $\tilde{t}(t)$ appearing in the various formulas of the $\Lambda$CDM-NG model. Finally, Eqs.(\ref{XFactor})-(\ref{PhysTime3}) determine the formula for the observed Hubble parameter,
\begin{equation}
H^{obs}(t)=H_{0}\sqrt{\Omega_{\Lambda}+\Omega_{m}\left(\frac{X}{a_{obs}}\right)^3+\Omega_{r}\left(\frac{X}{a_{obs}}\right)^4},
\label{HubObs3}
\end{equation}
which gives a possible resolution of the Hubble tension.

We conclude that the possible resolution of the Hubble tension offered by the $\Lambda$CDM-NG model is a result of the gravitational dilation of cosmic time; an effect which is not included in the $\Lambda$CDM model because, in that model, the average density of condensed matter is placed in the source term of the Friedman equation, where it does not produce potential $\bar{\Phi}$.

%*******************************
\section{Conclusion}
%******************************
The $\Lambda$CDM-NG model differs from the $\Lambda$CDM model in only one respect: The $\Lambda$CDM model places the average density of condensed masses $\bar{\rho}_{\Phi}$ in the source term of Friedmann's equation, whereas the $\Lambda$CDM-NG model places this density in the source term for the gravitational potential to restore this source term to its Newtonian form. That we have the freedom to do so within conventional general relativity, and without altering the accepted cosmological sources of gravitation, is the principle insight of this paper.

The Hubble tension is a feature of the $\Lambda$CDM model, and the $\Lambda$CDM-NG model avoids this tension if the gravitational potential of the universe, at the present time, falls in a certain range, that does not seem unreasonable for the actual universe, and does not require any fine tuning of parameters. A precision measurement of the deceleration parameter, to determine whether or not it has the value $q_{0}=(1/2)\Omega_{m}+\Omega_{r}-\Omega_{\Lambda}$ predicted by the $\Lambda$CDM model, could distinguish between the two models.

We close with three open questions: (1) Is Newton's theory of gravitation still valid on smaller scales in relativistic cosmology, or is the $\Lambda$CDM model strictly correct?; (2) Is the deceleration parameter determined by the density parameters as $q_{0}=(1/2)\Omega_{m}+\Omega_{r}-\Omega_{\Lambda}$ or not?, and (3) Is the Hubble tension the first quantitative evidence supporting the $\Lambda$CDM-NG model?


\begin{thebibliography}{99}

% Einstein's Initial Paper on General Relativity.   [1]

\bibitem[Einstein (1915)] {Einstein1} 
 Einstein, A. 1915, Preuss. Akad. Wiss. Berlin, Sitzber. 844-847; \textit{Annalen der Physik}, 49 (1916); Translation in: \textit{The Principle of Relativity}, (Dover Publications, Inc., 1952), pp. 112-164

% Einstein's First Paper on Cosmology [2]

\bibitem[Einstein (1917)] {Einstein2}
 Einstein, A. 1917, Sitzungsberichte der Preussischen Akad. d. Wissenschaften; Translation in: \textit{The Principle of Relativity}, (Dover Publications, Inc., 1952), pp. 177-188

%. Papers Showing Universe is Dynamic.  

%. [3]
\bibitem[De Sitter (1917a)]{De Sitter1}
 De Sitter, W. 1917, Proc. Kon. Ned. Akad. Wet. \textbf{19}, 1217-1225 
 
 % [4]
 \bibitem[De Sitter (1917b)]{De Sitter2}
 De Sitter, W.  1917, Proc. Kon. Ned. Akad. Wet. \textbf{20}, 229-243

%. [5]
\bibitem[Friedmann (1922)]{Friedmann1}
 Friedmann, A. 1922, Z. Phys. A \textbf{10} (1): 377-386

%. [6]
\bibitem[Friedmann (1924)]{Friedmann2}
 Friedmann, A. 1924, Z. Phys. A \textbf{21} (1): 326-332 

%. [7]
\bibitem[Lema\^itre (1931)]{Lemaitre1}
 Lema\^itre, G. 1931, Mon. Not. R. Astron. Soc. \textbf{91} (5): 483-490

%. [8]
\bibitem[Lema\^itre (1933)]{Lemaitre2}
Lema\^itre, G. 1933, Annales de la Soci\`et\`e Scientifique de Bruxelles,\textbf{A53:} 51-85
 
 %[9]
\bibitem[Robertson (1935)]{Robertson1}  Robertson, H.P. 1935,  Astrophys. J. \textbf{82}: 284-301

%. [10]
\bibitem[Robertson (1936a)]{Robertson2} Robertson, H.P. 1936a, Astrophys. J. \textbf{83}: 187-201

%. [11]
\bibitem[Robertson (1936b)]{Robertson3} Robertson, H.P.  1936b, Astrophys. J. \textbf{83}: 257-271

% [12]
\bibitem[Walker (1937)]{Walker}  Walker, A. G. 1937, Proceedings of the London Mathematical Society, series 2, \textbf{42} (1): 90-127

%. [13]
\bibitem[Eddington (1930)]{Eddington}  Eddington, A.S. 1930, Mon. Not. R. Astron. Soc., \textbf{78}: 3-28

% Dark Matter

%. [14]
 \bibitem[Zwicky (1933)]{Zwicky1} Zwicky, F. 1933, Helv. Phys. Acta. \textbf{6}: 110–127
  
%. [15]
\bibitem[Zwicky (1937)]{Zwicky2}  Zwicky, F. 1937,  Astrophys. J. \textbf{86}: 217–246

%. [16]
\bibitem[Rubin et al. (1980)]{Rubin}  Rubin, V., Thonnard, W.K.,  Ford, N. 1980, Astrophys. J. \textbf{238}: 471

%. [17]
\bibitem[Rogstad et al. (1972)]{Rogstad} Rogstad, D. H., Shostak, Seth, 1972,  Astrophys. J. \textbf{176}: 315–321

%. [18]
\bibitem[Perlmutter et at. (1999)]{Perlmutter} Perlmutter, S.,  Aldering, G., Goldhaber, G., et al.  1999,  Astrophys. J. \textbf{517} (2): 565–586, arXiv:astro-ph/9812133.

%. [19]
\bibitem[Riess (1998)]{Riess}  Riess, Adam G.,  Filippenko, Alexei V.,  Challis, Peter, et al. 1998,  Astron. J. \textbf{116} (3): 1009–1038,  arXiv:astro-ph/9805201.

%. General Lambda-CDM model

%. [20]
\bibitem[Weinberg (2008)]{Weinberg}  Weinberg, S. \textit{Cosmology} (Oxford University Press, Oxford, 2008).

%.  Cosmological Constant Problem

%. [21]
\bibitem[Weinberg (1989)] {Weinberg2}  Weinberg, S., 1989, Rev. Mod. Phys. \textbf{61} (1): 1-23

%. [22]
\bibitem[Adler (1995)]{Adler} Adler, R.,  Casey, B., Jacob, O. C. 1995, A. J. Phys. \textbf{63} (7): 620-626


%. James Web Early Galaxy Formation

%. [23]
\bibitem[Finkelstein et al. (2022)]{Finkelstein}  Finkelstein, S. L., Bagley, M. B.; Ferguson, H. C.; et. al., arXiv e-prints, arXiv:2211.05792. https://arxiv.org/abs/2211.05792.
%. [24]
\bibitem[Labbe et al. (2022)]{Labbe} Labbe, I., van Dokkum, P., Nelson, E., et. al. 2022, arXiv e-prints, arXiv:2207.12446. https://arxiv.org/abs/2207.12446

%. [25]
\bibitem[Naidu et al. (2022)]{Naidu}  Naidu, R. P., Oesch, P. A., Dokkum, P. V., et. al.  ApJ, 940, L14. doi: 10.3847/2041-8213/ac9b22


%. [26]
\bibitem[Adams et al. (2023)] {Adams}  Adams, N. J., Conselice, C.J., Ferreira, L., et. al. 2023, MNRAS, 518, 4755,  (2023). doi: 10.1093/mnras/stac3347

%. [27]
\bibitem[Rodighiero et al. (2011)]{Rodighiero}  Rodighiero, G., Daddi, E., Baronchelli, I., et. al. 2011, ApJ, 739, L40. doi: 10.1088/2041-8205/739/2/L40

%. Galaxy Formation

%. [28]
\bibitem[Haslbauer et al. (2022)] Haslbauer, M., Kroupa, P., Zonoozi, A. H., and  Haghi, H., 2022,  ApJ, 939, L31. doi: 10.3847/2041-8213/ac9a50.

%  [29]
\bibitem[Boylan-Kolchin (2022)]{Boylan-Kolchin}  Boylan-Kolchin, M., arXiv e-prints, arXiv:2208.01611. https://arxiv.org/abs/2208.01611.

%. Hubble Tension

%. [30]
\bibitem[Planck (2018)]{Planck} Planck Collaboration, 2018b, Planck 2018 results. VI. Cosmological parameters. arXiv:1807.06209

%. [31]
\bibitem[diValentino et al. (2021)]{diValentino}  di Valentino, Eleonora, et. al. 2021, Classical and Quantum Gravity \textbf{38} (15): 153001. arXiv:2103.01183 (https://arxiv.org/abs.2103.01183)

%. [32]
\bibitem[Bardeen (1980) pp.1882-1905]{Bardeen}  Bardeen, J.M. 1980, Phys. Rev. D \textbf{22}

%. [33]
\bibitem[Dodelson (2021) pp.143-147)]{Dodelson}  Dodelson, S. and  Schmidt, F. \textit{Modern Cosmology, 2nd Ed.}, (Academic Press, London, 2021)

%. [34]
\bibitem[Misner et al. (1973)]{Wheeler} Misner, C. W.; Thorn, K. S.; and  Wheeler, J. A.,\textit{Gravitation}, (W. H. Freeman and Company, San Francisco, 1973), Chap. 20.

%. [35]
\bibitem[Baumann (2022) pp.186-189]{Baumann}  Baumann, D., \textit{Cosmology}, (Cambridge University Press, New York, 2022), p. 224; .
 
%. [36]

\bibitem[Weinberg (1987)]{Weinberg3}  Weinberg, S. 1987, Phy. Rev. Lett. \textbf{59} 2607

%.  [37]  Sachs-Wolfe Effect

\bibitem[Sachs et al. (1967)]{Sachs}   Sachs, R. K., and  Wolfe, A. M.  1967,  Astrophys. J. \textbf{147}:73

%  Density Determinations Outside of Planck

%. [38]
\bibitem[Burles (1998)]{Burles} Burles, S., and  Tytler, D. 1998,  Astrophys. J. \textbf{499} 699-712

%. [39]
\bibitem[Cooke et al. (2018)]{Cooke}  Cooke, R. J., Pettini, M., Steidel, C.C. 2018, Astrophys. J. \textbf{855} (2), 102

%. [40]
\bibitem[Abbott (1918)]{Abbott}  Abbott, K. N., et al. 1918, Phys. Rev. D \textbf{98} (4) 043526

%. [41]
\bibitem[Mantz (2014)]{Mantz}  Mantz, A. B., et al. 2014, Mon. Not. R. Astron. Soc. \textbf{440} (3), 2077-2098



\end{thebibliography}
\end{document}